\documentclass[aps,pre,twocolumn,superscriptaddress,nofootinbib,floatfix]{revtex4-2}

\usepackage[utf8]{inputenc}
\usepackage[T1]{fontenc}
\usepackage{amsmath,amssymb,mathtools}
\usepackage{graphicx}
\usepackage{booktabs}
\usepackage{hyperref}

\hypersetup{
    colorlinks=true,
    linkcolor=blue,
    citecolor=blue,
    urlcolor=blue
}

\usepackage{tikz}
\usetikzlibrary{calc}
\usetikzlibrary{arrows.meta, positioning, shapes.geometric}

\begin{document}

\title{Neural-network solution of subtracted three-body Faddeev integral equations near the Efimov limit}

\author{Lucas A. Souza}
\email{lasouza@if.usp.br}
\affiliation{Minas Gerais, Brazil}

\date{\today}

\begin{abstract}
 
We apply a deep-neural-network (DNN) ansatz to the symmetrized spectator vector of the subtracted three-body Faddeev integral equation 
for identical bosons near the Efimov limit. The network is trained by minimizing the residual of the discretized integral equation, 
while the positive binding scale associated with the three-body energy is treated as a trainable parameter. Deterministic diagonalization of the 
same discretized kernel is used only as an a posteriori numerical benchmark. As preliminary validation, the neural-solver strategy is 
tested on the analytically solvable hydrogen radial problem. At unitarity, the DNN reproduces the Efimov ground-state binding scale with 
a DNN--deterministic deviation of $0.022\%$, while the first excited state is recovered to $0.002\%$. The deterministic solver recovers 
the universal Efimov scaling ratio $e^{2\pi/s_0}\simeq 515.03$, and the neural method traces the bound-state branches as a function of 
the inverse scattering length $1/a$ by continuation from the unitary solution. These results indicate that DNN-based residual minimization 
can provide a compact and differentiable representation of a renormalized few-body integral-equation solution in a regime governed by 
discrete scale invariance.

\end{abstract}

\maketitle

\section{Introduction}
\label{sec:introduction}
 
Deep neural networks are increasingly used in scientific computing as trainable representations of functions, fields, and wave functions~\cite{LeCunBengioHinton2015DeepLearning}.
A classical early example in scientific computing is the work of Lagaris \emph{et al.}, 
who embedded feed-forward neural networks in trial solutions satisfying boundary or initial 
conditions by construction and optimized the remaining parameters through the differential-equation 
residual~\cite{Lagaris1998ANN}. This idea was later generalized into physics-informed neural networks (PINNs) 
for forward and inverse problems governed by partial differential equations~\cite{Raissi2019PINN,Karniadakis2021PIML}.  
In computational quantum mechanics, this viewpoint is naturally connected to the use of neural networks as variational 
representations of wave functions or eigenstates. Most applications, however, are formulated in terms of differential equations 
or variational energy functionals, where the network represents a wave function in coordinate or momentum space and the loss is 
built from a Schrödinger operator, a Rayleigh quotient, 
or a variational Monte Carlo objective~\cite{CarleoTroyer2017NQS,Hermann2020PauliNet,Sarkar2025QuantumWellPINN,Brevi2024AnharmonicPINN}.

Nuclear and few-body physics provide a useful testing ground for neural-network solvers. 
Neural-network representations have been applied to the deuteron and to few-body nuclear systems, including studies of architecture dependence, 
uncertainty quantification, and bound-state calculations~\cite{KeebleRios2020DeuteronML,RozalenSarmiento2024DeuteronUQ,Brevi2026DeuteronPINN}. 
In particular, Li \emph{et al.} recently used DNNs for deuteron and triton calculations with realistic nuclear interactions~\cite{Li2026DNNFewBody}, 
providing a close few-body benchmark for neural wave-function methods. 
Recent work has also extended PINN-based solvers to nuclear scattering through exterior complex scaling, making oscillatory scattering boundary 
conditions compatible with neural-network representations~\cite{lei2026exteriorcomplexscalingenables}. 
More broadly, machine-learning methods are increasingly used in nuclear theory for emulation, inference, classification, uncertainty quantification, and many-body wave-function representations~\cite{Boehnlein2022MLNuclear}.
In nuclear astrophysics, Bayesian and neural-network strategies have also been used to constrain neutron-star equations of state from observational data~\cite{Chimanski2023,Baker2026}. 
Nevertheless, most existing quantum applications, including the few-body examples above, rely on differential, variational, or wave-function-based formulations. Direct neural-network treatments of integral-equation formulations, which are central to few-body scattering and bound-state theory, remain less explored.

A natural and demanding benchmark for an integral-equation neural solver is the three-boson system near the unitary limit. 
In this regime, the Efimov effect produces a geometric spectrum of three-body bound states governed by discrete scale invariance~\cite{efimov1970,efimov1973,braaten2006,naidon2017}. 
This universal physics has been extensively discussed in cold atoms and in few-body effective field theory, 
with applications ranging from atomic gases to nuclear systems~\cite{BraatenHammer2007ColdAtomsEfimov,HammerPlatter2010EfimovNuclearParticle,Kievsky2021EfimovNuclearConnections,HammerHiga2008DiscreteScaleInvariance}, and it also 
underlies universal descriptions of halo nuclei and halo effective field theory near the 
unitary limit~\cite{Kievsky2021EfimovNuclearConnections,Amorim1997UniversalHaloEfimov,Yamashita2008Neutron19C,Ji2025,Souza2016}. 
For zero-range interactions, however, the three-body equation is ultraviolet divergent: 
the Thomas collapse reflects the need for an additional three-body scale~\cite{Adhikari1988EfimovThomasModelDependence}. 
Subtracted Faddeev or Skorniakov--Ter-Martirosian formulations introduce this scale by replacing the divergent 
kernel with a kernel subtracted at a fixed three-body energy, thereby trading cutoff dependence for 
a renormalization scale~\cite{Frederico1999Subtracted}.

This structure makes the problem well suited for testing equation-informed neural 
representations beyond standard differential-equation benchmarks. The unknown object is not 
a many-body wave function in full configuration space, but a single-variable spectator function 
satisfying a homogeneous integral equation. The binding energy enters nonlinearly through both the 
kernel and the two-body prefactor, so the bound-state condition is an eigenvalue condition rather than 
a prescribed-energy residual. Recent neural-network studies have also addressed Efimov and few-body physics using 
wave-function-based neural quantum states or few-body machine-learning frameworks~\cite{Saito2018FewBodyANN,Yokoi2026EfimovNN,Ziqi2026FewBodyML}. 
The present work follows a different route: the DNN represents the symmetrized spectator vector of the subtracted integral equation itself, 
and the loss is the residual of the discretized Faddeev kernel. 

In this work we develop a DNN ansatz for the subtracted three-body spectator equation of identical bosons near the Efimov limit. 
The three-body binding scale $\epsilon_3=-E_3>0$ is treated as a trainable parameter, optimized jointly with the network weights, 
so that the neural calculation does not require an external spectral scan during training. A deterministic diagonalization of 
the same discretized kernel is used only as an a posteriori numerical benchmark. We first validate the DNN training pipeline 
on the analytically solvable hydrogen radial problem, where exact energies and wave functions are available. We then apply 
the method to the Efimov problem at unitarity and along the ground-state branch as a function of the inverse scattering 
length $1/a$. The goal is to demonstrate that residual minimization with a trainable binding scale can construct a compact, 
differentiable representation of a renormalized few-body integral-equation solution.

The paper is organized as follows. Section~\ref{sec:faddeev} introduces the subtracted three-body 
spectator equation and its deterministic discretization. Section~\ref{sec:neural_method} describes 
the DNN representation, the trainable binding-energy parameter, and the residual loss. Section~\ref{sec:benchmark_basic} 
presents the hydrogen benchmark. Section~\ref{sec:results_efimov} gives the Efimov-regime results, including the unitary 
spectrum, the universal ratios, the DNN--deterministic deviations, and the continuation in $1/a$. Section~\ref{sec:discussion} 
discusses the computational meaning, limitations, and possible extensions of the method.

\section{Subtracted three-body Faddeev integral equation}
\label{sec:faddeev}
  
\subsection{Zero-range interaction, spectator function, and subtraction}

We consider three identical bosons interacting through a short-range $s$-wave force in the zero-range limit. Formally, this limit may be represented by a contact interaction,
\[
V(\mathbf r)=\lambda\,\delta^{(3)}(\mathbf r),
\]
but in three spatial dimensions such an interaction is not a regular potential: it must be understood as a renormalized contact interaction specified by the two-body scattering length $a$. The two-body subsystem has a shallow bound state for $a>0$, with
\[
E_2=-\epsilon_2,\qquad \epsilon_2=\frac{1}{a^2},
\]
in the units used below. For $a<0$, the same zero-range amplitude describes a virtual-state pole through the sign of $1/a$. The unitary limit corresponds to $1/a=0$.

For a bound state we write the physical three-body energy as
\(E_3=-\epsilon_3\), with \(\epsilon_3>0\). In these units, the free three-body denominator is
\(\epsilon_3+p^2+3q^2/4\), up to an overall sign absorbed in the normalization. For three identical bosons, the momentum-space wave function has the schematic form
\begin{equation}
\Psi(\mathbf p,\mathbf q)
\propto
\frac{
f(q)
+
f(|\mathbf p-\mathbf q/2|)
+
f(|\mathbf p+\mathbf q/2|)
}{
\epsilon_3+p^2+3q^2/4
}.
\label{eq:wavefunction_spectator}
\end{equation}
Here \(f(q)\) is the spectator amplitude, \(q\) is the spectator momentum, and \(\mathbf p\) is the relative momentum inside the interacting pair.

The corresponding zero-range Skorniakov--Ter-Martirosian equation~\cite{skorniakov1957} is ultraviolet divergent. 
This is the momentum-space manifestation of the Thomas collapse~\cite{thomas1935}: without an additional three-body input, 
the spectrum is not bounded from below as the ultraviolet cutoff is removed. Equivalently, the Efimov effect~\cite{efimov1970} 
and the Thomas collapse are two aspects of the same singular structure of the zero-range three-body kernel~\cite{Frederico1999Subtracted,frederico2012, braaten2006}. 
A three-body scale is therefore required to define the problem.

We introduce this scale through a subtraction at the energy $-\mu_3^2$. The physical kernel at energy $-\epsilon_3$ is replaced by the difference between this kernel and the same kernel evaluated at the subtraction scale. In this way, the ultraviolet dependence is traded for the finite three-body scale $\mu_3$, which fixes the absolute position of the Efimov spectrum. Throughout this work we set $\mu_3=1$ and express all momenta and energies in units of this scale.

The resulting subtracted spectator equation is
\begin{equation}
\begin{aligned}
f(q)
&=
\frac{2/\pi}
{\sqrt{\epsilon_3+3q^2/4}-1/a}
\int_0^\infty dq'\,
\frac{q'}{q}
\\
&\qquad\times
\left[
L_{\epsilon}(q,q')-L_{\mu}(q,q')
\right] f(q') .
\end{aligned}
\label{eq:spectator}
\end{equation}
with
\begin{align}
L_{\epsilon}(q,q')
&=
\ln
\left(
\frac{
\epsilon_3+q^2+q'^2+qq'
}{
\epsilon_3+q^2+q'^2-qq'
}
\right),
\label{eq:Leps}
\\
L_{\mu}(q,q')
&=
\ln
\left(
\frac{
\mu_3^2+q^2+q'^2+qq'
}{
\mu_3^2+q^2+q'^2-qq'
}
\right).
\label{eq:Lmu}
\end{align}
The subtraction term $L_\mu$ removes the leading ultraviolet behavior of the zero-range kernel. The difference $L_\epsilon-L_\mu$ is therefore the central object of the renormalized formulation: changing $\mu_3$ fixes the three-body parameter and shifts the absolute spectrum, while the ratios between sufficiently shallow Efimov states approach the universal discrete-scaling factor.

\subsection{Discretization and deterministic benchmark}

To solve Eq.~\eqref{eq:spectator} numerically, we discretize the momentum variable $q$ on a grid $\{q_i\}_{i=1}^{N}$ with quadrature weights $\{w_i\}$. The integral equation becomes a matrix eigenvalue problem
\begin{equation}\label{eq:matrix}
    \sum_{j=1}^{N} K_{ij}(\epsilon_3, 1/a)\,f_j = \eta\,f_i,
\end{equation}
where the non-symmetric kernel matrix has elements
\begin{multline}
    K_{ij} = w_j\,
             \frac{2}{\pi\bigl(\sqrt{\epsilon_3 + 3q_i^2/4} - 1/a\bigr)}\,
             \frac{q_j}{q_i} \\[4pt]
             \times \bigl[L_\epsilon(q_i,q_j) - L_\mu(q_i,q_j)\bigr].
\end{multline}
Physically acceptable bound states correspond to eigenvalues $\eta = 1$: only when the kernel has a unit eigenvalue does the homogeneous equation admit a nontrivial solution. The three-body energy $\epsilon_3$ is therefore determined implicitly by the condition $\eta_{\max}(\epsilon_3) = 1$, where $\eta_{\max}$ is the largest eigenvalue of $K$.

For stable numerical diagonalization, we symmetrize Eq.~\eqref{eq:matrix} by introducing the rescaled vector
\begin{equation}
    u_i \propto \frac{\sqrt{w_i}\,q_i f_i}{\sqrt{P_i}},
    \qquad
    P_i = \frac{2}{\pi\bigl(\sqrt{\epsilon_3 + 3q_i^2/4} - 1/a\bigr)},
\end{equation}
where $P_i$ is the two-body prefactor evaluated at grid point $i$. With this definition, the eigenvalue problem takes the symmetric form
\begin{multline}\label{eq:sym}
    u = B(\epsilon_3, 1/a)\,u, \\[4pt]
    B_{ij} = \sqrt{w_i P_i\,w_j P_j}\;
             \bigl[L_\epsilon(q_i,q_j) - L_\mu(q_i,q_j)\bigr].
\end{multline}
The original spectator function is recovered through $f_i \propto \sqrt{P_i}\,u_i / (\sqrt{w_i}\,q_i)$.
The matrix $B$ is real symmetric by construction; its eigenvalues are real and its eigenvectors are orthogonal. The deterministic reference solution --- our numerical benchmark --- is obtained by scanning $\epsilon_3$ on a fine logarithmic grid and bisecting the crossing condition $\eta_{\max}(\epsilon_3) = 1$ for the largest eigenvalue of $B$. This procedure yields both the reference energy $\epsilon_3^{\rm(ref)}$ and the reference eigenvector $u^{\rm(ref)}$, against which the DNN results are compared.

To resolve the wide range of momentum scales required by the Efimov spectrum --- from $\epsilon_3 \simeq 10^{-2}$ for the deepest state shown to $\epsilon_3 \simeq 10^{-10}$ for the shallowest branch --- we employ Gauss--Legendre quadrature on a logarithmic grid,
\begin{equation}
    q = e^t,\qquad t \in [\log q_{\min}, \log q_{\max}],
\end{equation}
with the Jacobian $dq = q\,dt$ absorbed into the quadrature weights. The logarithmic spacing concentrates points at small $q$, where shallow bound-state wave functions are localized. Throughout this work we use $N_q = 600$, $q_{\max}/\mu_3 = 4$, and $q_{\min}/\mu_3 = 10^{-12}$.

\section{Neural-network representation and loss function}
\label{sec:neural_method}

\begin{figure*}
\centering
\begin{tikzpicture}[
    node distance=0.4cm and 1.4cm,
    >=Latex,
    every node/.style={font=\footnotesize},
    box/.style={
        draw,
        fill=gray!10,
        rounded corners,
        minimum width=2.0cm,
        minimum height=0.9cm,
        align=center
    },
    smallbox/.style={
        draw,
        fill=gray!10,
        rounded corners,
        minimum width=1.8cm,
        minimum height=0.8cm,
        align=center
    },
    circ/.style={
        draw,
        fill=gray!10,
        circle,
        minimum size=0.7cm,
        align=center
    }
]

\node[circ] (input) {$x = \log(q/\mu_3)$};
\node[box, right=of input] (dnn) {Deep neural\\network};
\node[box, right=of dnn] (output) {Symmetrized spectator\\vector $u_\theta(x)$};

\node[smallbox, below=0.5cm of output] (res) {Residual of the\\integral equation};
\node[smallbox, left=of res] (norm) {Normalization};
\node[smallbox, right=of res] (energy) {Trainable energy\\$\epsilon_3 = \exp(s_3)$};
\node[smallbox, below=of res] (loss) {Total loss};

\node[smallbox, left=of norm] (orth) {Orthogonality\\(if applicable)};

\draw[->] (input) -- (dnn);
\draw[->] (dnn) -- (output);

\draw[->] (output.south) -- (res.north);
\draw[->] (energy.west) -- (res.east);

\draw[->] (res.south) -- (loss.north);
\draw[->] (norm.south) -- (loss.north west);
\draw[->] (orth.south east) -- (loss.west);
\draw[->] (energy.south) |- (loss.east);

\draw[->, dashed] (loss.west) .. controls +(-4.0,0) and +(0,-3.0) .. ($(dnn.south)!0.5!(dnn.south west)$);

\end{tikzpicture}
\caption{
Schematic representation of the DNN solver for the subtracted Faddeev integral equation.
The logarithmic momentum $x = \log(q/\mu_3)$ is fed into a feed-forward neural network,
which outputs the symmetrized spectator vector $u_\theta(x)$.
The binding energy $\epsilon_3 = \exp(s_3)$, with $s_3 = \log\epsilon_3$, is a trainable
parameter and enters the residual of the discretized integral equation.
The total loss combines the equation-informed residual with normalization and,
when required, orthogonality constraints.
The dashed feedback arrow indicates that gradients flow back through the kernel
$B(\epsilon_3, 1/a)$ to both the network weights $\theta$ and the energy parameter $s_3$.
}
\label{fig:dnn_schematic}
\end{figure*}
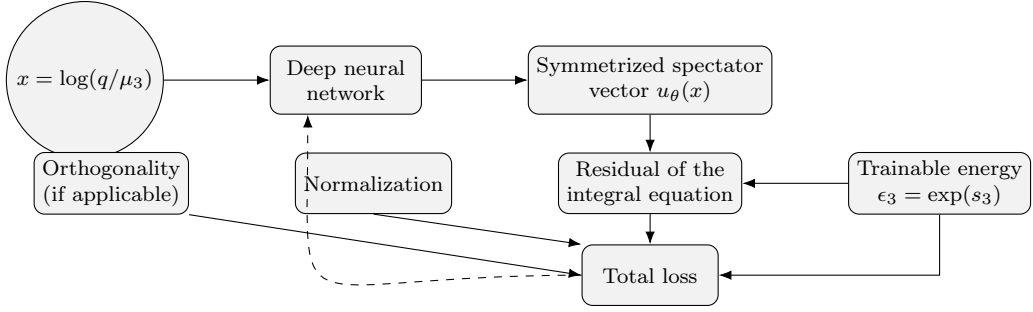

The DNN implementation is summarized in Fig.~\ref{fig:dnn_schematic}. We represent the symmetrized spectator vector $u$ by a feed-forward neural network $u_\theta$ whose input is the logarithmic momentum variable $x = \log(q/\mu_3)$. The network does not represent the original spectator function $f(q)$ directly; the two are related by the symmetrization transformation $f_i \propto \sqrt{P_i}\,u_i/(\sqrt{w_i}\,q_i)$, as described in Sec.~\ref{sec:faddeev}. The architecture consists of $L = 3$ hidden layers with $H = 64$ neurons each and $\tanh$ activation, followed by a linear output layer, totaling $N_{\rm params} \approx 8\,500$ trainable parameters. All computations are performed in double precision (\texttt{float64}). The raw network output $\tilde{u}_\theta(x)$ is explicitly normalized before evaluating any physical quantity:
\begin{equation}
    u_\theta(q_i) = \frac{\tilde{u}_\theta(x_i)}
                         {\sqrt{\sum_j \tilde{u}_\theta(x_j)^2 + \delta_{\rm num}}},
\end{equation}
with $\delta_{\rm num} = 10^{-30}$. This normalization removes the scale ambiguity of the homogeneous equation and ensures that the loss function measures only the shape mismatch of the spectator function.

The three-body binding energy is treated as a trainable parameter. To enforce $\epsilon_3 > 0$ without constrained optimization, we parametrize $\epsilon_3 = \exp(s_3)$, where $s_3 = \log\epsilon_3$ is an unconstrained trainable scalar, and optimize $s_3$ jointly with the network weights $\theta$. The kernel matrix $B(\epsilon_3, 1/a)$ is rebuilt at each training epoch because $\epsilon_3$ changes.

The network is trained by minimizing the equation-informed residual of the discretized integral equation. The residual loss is defined as
\begin{equation}\label{eq:loss_res}
    \mathcal{L}_{\rm res}(\theta, \epsilon_3) =
    \frac{\sum_i \bigl(u_i - [B(\epsilon_3, 1/a)\,u]_i\bigr)^2}
         {\sum_i u_i^2},
\end{equation}
which is the normalized squared norm of the mismatch between $u$ and $B u$. The denominator makes $\mathcal{L}_{\rm res}$ independent of the overall scale of $u$, so that the normalization step does not interfere with the gradient of the residual. A weak regularizer on the raw network output,
\begin{equation}
    \mathcal{L}_{\rm norm} = \Bigl(\sum_i \tilde{u}_i^2 - 1\Bigr)^2,
\end{equation}
encourages the raw output to remain near unit norm, improving numerical stability. The total loss is
\begin{equation}
    \mathcal{L} = \mathcal{L}_{\rm res} + \lambda_{\rm norm}\mathcal{L}_{\rm norm},
\end{equation}
with $\lambda_{\rm norm} = 10^{-3}$. For an exact eigenvector of the discretized problem, $u = B u$ and the residual is identically zero. In practice, the DNN drives $\mathcal{L}_{\rm res}$ toward a small finite value, limited by optimization accuracy and quadrature resolution; typical converged residuals lie in the range $10^{-4}$--$10^{-5}$.

The optimizer is Adam~\cite{KingmaBa2015Adam} with separate learning rates: $\eta_\theta = 3\times10^{-3}$ for the network weights and $\eta_\epsilon = 5\times10^{-4}$ for the energy parameter. The smaller learning rate for $s_3$ prevents the energy from overshooting during early epochs, when the spectator function is far from converged. At each value of $1/a$, training proceeds for a fixed number of epochs, and the best model state is tracked by the minimum residual loss.

To trace the ground-state energy $E_3(1/a)$, we adopt a continuation strategy. Starting from the converged solution at unitarity ($1/a = 0$), the network weights and $s_3$ are initialized from the solution at the previous value of $1/a$ and trained for $N_{\rm cont}$ epochs. The continuation grid covers $1/a \in [-6.5\times10^{-2},\, 5.0\times10^{-1}]$ with points distributed logarithmically in $|1/a|$, yielding higher density near unitarity where the bound-state energies vary most rapidly. The negative and positive sides use the same logarithmic spacing in $|1/a|$. On the positive-$1/a$ side, the two-body prefactor in Eq.~\eqref{eq:spectator} develops a pole at the atom-dimer threshold $\epsilon_3 = (1/a)^2$; training points must satisfy $\epsilon_3^{\rm (DNN)} < (1/a)^2$ to remain below this threshold. Points for which the converged DNN energy would violate this condition are excluded from the continuation.

All DNN results shown in the figures and tables are obtained from a single continuation run starting at unitarity, with a fixed random seed of $1234$ for the network initialization and the Adam optimizer. The energies reported in Tables~III and IV correspond to this single run; no averaging or selection across multiple seeds was performed. A limited sensitivity study varying the seed was conducted and produced ground-state energies consistent to within $0.02\%$ at unitarity. No data labels, pretrained solutions, or supervised targets are used at any stage of training. The deterministic benchmark is computed entirely independently of the DNN and serves only for a posteriori comparison. When the overlap between the DNN and deterministic spectator amplitudes is evaluated, it is computed directly in the symmetrized space as $\langle u^{\rm (DNN)} | u^{\rm (det)} \rangle$, which is the natural inner product for the symmetrized eigenproblem. At unitarity, the ground-state DNN--deterministic overlap exceeds $0.9999$. The best model state at each $1/a$ is selected as the one achieving the minimum residual loss during training; no early stopping or validation set is employed, since the problem has no train/test split in the usual supervised-learning sense. The training history and final parameters for all runs are included in the supplemental material. The code and data used to produce the results of this work will be made publicly available upon publication.

\begin{table}
    \centering
    \caption{Neural-network and quadrature parameters used throughout this work.}
    \label{tab:params}
    \begin{tabular}{@{}ll@{}}
        \toprule
        Parameter & Value \\
        \midrule
        Hidden layers $L$ & 3 \\
        Hidden width $H$ & 64 \\
        Activation & $\tanh$ \\
        Trainable parameters & $\approx 8\,500$ \\
        Optimizer & Adam \\
        $\eta_\theta$ (network) & $3\times10^{-3}$ \\
        $\eta_\epsilon$ (energy) & $5\times10^{-4}$ \\
        $\lambda_{\rm norm}$ & $10^{-3}$ \\
        $\delta_{\rm num}$ & $10^{-30}$ \\
        $N_q$ & 600 \\
        $q_{\max}/\mu_3$ & 4 \\
        $q_{\min}/\mu_3$ & $10^{-12}$ \\
        Grid type & logarithmic \\
        Epochs (first point) & 5000 \\
        Epochs (continuation) & 400--5000 (adaptive) \\
        Random seed & 1234 \\
        \bottomrule
    \end{tabular}
\end{table}

\section{Deep-learning benchmark: hydrogen atom}
\label{sec:benchmark_basic}

Before addressing the subtracted three-body integral equation, we validate the neural-solver strategy on a quantum system with known analytical solutions. The purpose of this benchmark is not to test the Faddeev kernel itself, but to verify that the DNN implementation can reproduce bound-state energies and wave functions in a controlled eigenvalue problem.

We consider the reduced radial Coulomb problem for the electron--proton system. 
In dimensionless units, with
\[
\rho=\frac{r}{a_\mu},
\qquad
\mu=\frac{m_e m_p}{m_e+m_p},
\]
where \(a_\mu\) is the Bohr radius defined with the reduced mass \(\mu\), the reduced radial wave function \(u(\rho)\) satisfies
\begin{equation}
\left[
-\frac{1}{2}\frac{d^2}{d\rho^2}
+
\frac{\ell(\ell+1)}{2\rho^2}
-\frac{1}{\rho}
\right]u(\rho)
=
E u(\rho).
\label{eq:hydrogen_radial}
\end{equation}
The exact spectrum is $E_n=-1/(2n^2)$ in Hartree atomic units. In the comparison below, energies are reported in these dimensionless units.

For this benchmark, the DNN is used in configuration space rather than in momentum space. The network input is the radial coordinate $\rho$, and the output is the reduced radial wave function $u_\theta(\rho)$. This differs from the Efimov calculation, where the input is the logarithmic momentum variable $x=\log(q/\mu_3)$ and the network represents the symmetrized spectator vector. The common element is the numerical strategy: a DNN is used as a differentiable representation of the unknown function and is optimized through an equation-based objective.

The hydrogen calculation is formulated variationally through the Rayleigh quotient. For a given angular momentum $\ell$, the DNN energy is
\begin{equation}
E_\theta^{(\ell)}
=
\frac{
\displaystyle
\int_0^{\rho_{\max}}
d\rho\,
\left[
\frac{1}{2}
\left(
\frac{d u_\theta}{d\rho}
\right)^2
+
\left(
\frac{\ell(\ell+1)}{2\rho^2}
-\frac{1}{\rho}
\right)
u_\theta^2
\right]
}{
\displaystyle
\int_0^{\rho_{\max}}
d\rho\,u_\theta^2
}.
\label{eq:hydrogen_rayleigh}
\end{equation}
In the numerical implementation, the integrals are evaluated by Gauss--Legendre quadrature on the finite interval $[0,\rho_{\max}]$. 
The reduced radial formulation requires regular solutions at the origin, $u(0)=0$, while the finite radial box is chosen large enough that the bound-state wave functions have reached their asymptotic tail.
 
For the hydrogen benchmark, the variational objective is based on the Rayleigh quotient in Eq.~\eqref{eq:hydrogen_rayleigh}. 
Since the Rayleigh quotient is invariant under an overall rescaling of the trial function, the network output is normalized on the quadrature grid when evaluating energies and overlaps. 
For excited states in the same angular-momentum sector, an orthogonality constraint is imposed with respect to the previously obtained lower state. 
In particular, the $2s$ state is constrained to be orthogonal to the $1s$ reference state through
\begin{equation}
\mathcal{L}_{\rm orth}^{(2s)}
=
\left|
\int_0^{\rho_{\max}}
d\rho\,
u_\theta^{(2s)}(\rho)
u_{\rm ref}^{(1s)}(\rho)
\right|^2 .
\label{eq:hydrogen_orth}
\end{equation}
Thus, the hydrogen training objective is based on the Rayleigh quotient, with normalization of the trial function and, when required, an orthogonality penalty. 
The $2p$ state is obtained by solving the radial 
equation in the $\ell=1$ sector; it therefore tests the centrifugal term and belongs to a different angular sector from the $s$ states.
  
We study the $1s$, $2s$, and $2p$ states. This set provides a compact validation of the neural solver: the $1s$ state tests the ground-state variational solution, the $2s$ state tests the representation of a radial node and the orthogonality constraint, and the $2p$ state tests the treatment of the centrifugal term. The exact reduced radial wave functions used for comparison are
\begin{align}
u_{1s}(\rho) &\propto \rho e^{-\rho},\\
u_{2s}(\rho) &\propto \rho(2-\rho)e^{-\rho/2},\\
u_{2p}(\rho) &\propto \rho^2 e^{-\rho/2}.
\end{align}

\begin{figure}
    \centering
    \includegraphics[width=\columnwidth]{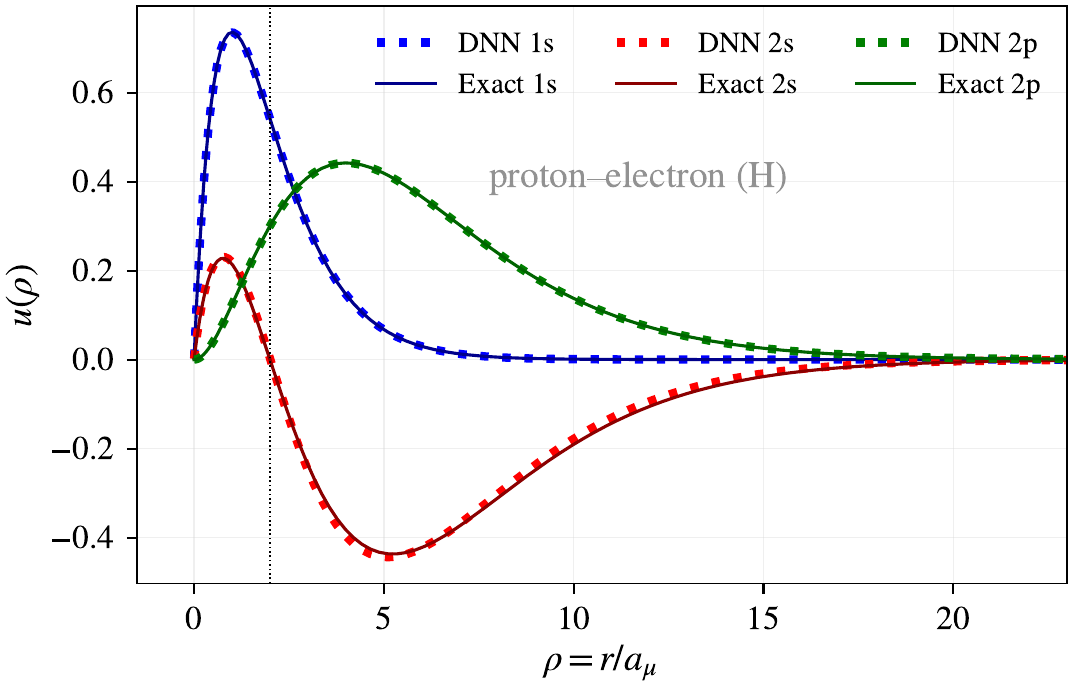}
    \caption{The DNN curves closely follow the analytical results; the $2s$ node at $\rho=2$ is correctly captured.}
    \label{fig:hydrogen_radial}
\end{figure}

The resulting radial wave functions are shown in Fig.~\ref{fig:hydrogen_radial}. The DNN solutions reproduce the exact analytical curves for all three states, including the radial node of the $2s$ state at $\rho=2$ and the broader radial profile of the $2p$ state. The corresponding energies and overlaps are summarized in Table~\ref{tab:hydrogen}. The largest energy deviation occurs for the nodal $2s$ state, with a relative error of $2.17\times10^{-3}$, while the overlap remains $0.999274$. The $1s$ and $2p$ states are reproduced to substantially higher accuracy. This benchmark confirms that the DNN training pipeline can reproduce both eigenvalues and wave functions in a problem where exact analytical results are available, before applying the method to the subtracted three-body integral equation.

\begin{table}
    \centering
    \caption{
    Hydrogen benchmark: DNN results compared with the exact analytical spectrum.
    Energies are given in Hartree atomic units. The relative error is defined as
    $|E_{\rm DNN}-E_{\rm exact}|/|E_{\rm exact}|$.
    }
    \label{tab:hydrogen}
    \begin{tabular}{@{}lccccc@{}}
        \toprule
        State & $\ell$ & $E_{\rm exact}$ & $E_{\rm DNN}$ & Rel.\ error & Overlap \\
        \midrule
        $1s$ & 0 & $-0.500000$ & $-0.499998$ & $4.30\times10^{-6}$ & $1.000000$ \\
        $2s$ & 0 & $-0.125000$ & $-0.124729$ & $2.17\times10^{-3}$ & $0.999274$ \\
        $2p$ & 1 & $-0.125000$ & $-0.125000$ & $2.98\times10^{-7}$ & $1.000000$ \\
        \bottomrule
    \end{tabular}
\end{table}

\section{Results: Efimov regime}
\label{sec:results_efimov} 

We first validate that the discretized subtracted kernel correctly reproduces the Efimov spectrum. 
Using $\mu_3 = 1$, $q_{\max}/\mu_3 = 4$, and $N_q = 600$ on a logarithmic grid, we diagonalize the symmetrized matrix $B(\epsilon_3)$ of 
Eq.~\eqref{eq:sym} and locate the bound states by bisecting the crossing condition $\eta_{\max}(\epsilon_3) = 1$. At unitarity ($1/a = 0$), 
this yields the first four three-body bound states, whose positive energy parameters $\epsilon_3$ are listed in the first column of Table~\ref{tab:dnn_vs_det}.

The ratios between successive states provide a stringent test of the discretization. The deterministic calculation gives
\begin{equation}
    \frac{\epsilon_3^{(0)}}{\epsilon_3^{(1)}} \simeq 507.75,\qquad
    \frac{\epsilon_3^{(1)}}{\epsilon_3^{(2)}} \simeq 515.02,\qquad
    \frac{\epsilon_3^{(2)}}{\epsilon_3^{(3)}} \simeq 515.03.
\end{equation}
The ratios for the two shallowest pairs reproduce the universal Efimov factor $e^{2\pi/s_0} \simeq 515.04$ to within $2\times10^{-5}$. 
The ground-to-first-excited ratio ($507.75$) lies $\approx 1.4\%$ below the universal value: this is a non-asymptotic correction arising 
because the deepest state probes the short-distance scale $\mu_3$, which here acts as the renormalization point. 
The recovery of the universal ratios confirms that the discretized subtracted kernel captures the Efimov discrete scaling. 
Numerical convergence with respect to $N_q$ and $q_{\max}$ was verified separately by varying the quadrature grid parameters. 
The shallow branches span roughly eight orders of magnitude in $\epsilon_3$ (from $\sim 10^{-2}$ to $\sim 10^{-10}$); 
the logarithmic grid with $N_q = 600$ provides sufficient resolution across this range.

Having established the deterministic benchmark, we apply the DNN solver with trainable energy to the same system at unitarity. 
Table~\ref{tab:dnn_vs_det} compares the DNN-predicted energies against the deterministic values. For the ground state ($n=0$), 
the DNN yields $\epsilon_3^{\rm (DNN)} = 9.2719\times10^{-3}$, deviating from the benchmark by merely $0.022\%$. 
The first excited state is reproduced with a deviation of $0.002\%$. For the shallower branches $n=2$ and $n=3$, 
the DNN--deterministic deviations grow to $2.890\%$ and $3.359\%$, respectively. 
This trend is expected: the spectrum compresses geometrically, $\epsilon_3^{(n)}/\epsilon_3^{(n+1)} \sim 515$, 
so the absolute energy scale of each successive branch is roughly three orders of magnitude smaller. 
The more weakly bound states are therefore exponentially more sensitive to any residual mismatch in the spectator function, 
and the loss landscape becomes flatter near the solution.

Table~\ref{tab:efimov_ratios} compares the Efimov ratios extracted from both methods. The DNN ratios are $507.87$, $500.54$, and $512.70$ for the successive pairs. The deviation of the DNN ratio from $e^{2\pi/s_0}$ is $-0.45\%$ for the shallowest pair ($n=2,3$), showing that the network captures the universal discrete scaling despite the $\sim 3\%$ deviation in the individual energies of branch $n=3$. For the intermediate pair ($n=1,2$), the deviation reaches $-2.81\%$; this is the pair most affected by the accumulated uncertainty in branch $n=2$. The ground-to-first-excited ratio is the least sensitive to the DNN approximation because both states are relatively deep and well resolved.

\begin{table}
    \centering
    \caption{%
        Comparison between the deterministic solution and the DNN-predicted
        three-body bound-state energies at unitarity ($1/a = 0$).
        $\epsilon_3 = -E_3 > 0$ is the positive binding-energy parameter.
        The DNN results are obtained with the trainable-energy scheme
        (Sec.~\ref{sec:neural_method}).
    }
    \label{tab:dnn_vs_det}
    \begin{tabular}{@{}c c c c@{}}
        \toprule
        $n$ &
        $\epsilon_3^{\rm (det)}$ &
        $\epsilon_3^{\rm (DNN)}$ &
        DNN--det.\ deviation (\%) \\
        \midrule
        0 & $9.2699 \times 10^{-3}$ & $9.2719 \times 10^{-3}$ & 0.022 \\
        1 & $1.8257 \times 10^{-5}$ & $1.8256 \times 10^{-5}$ & 0.002 \\
        2 & $3.5449 \times 10^{-8}$ & $3.6473 \times 10^{-8}$ & 2.890 \\
        3 & $6.8828 \times 10^{-11}$ & $7.1140 \times 10^{-11}$ & 3.359 \\
        \bottomrule
    \end{tabular}
\end{table}

\begin{table}
    \centering
    \caption{%
        Ratios $\epsilon_3^{(n)} / \epsilon_3^{(n+1)}$ of successive
        three-body bound states at unitarity, comparing the deterministic
        solution with the DNN prediction.
        The last column gives the relative deviation of the DNN ratio
        from the universal Efimov factor $e^{2\pi/s_0} \simeq 515.04$.
    }
    \label{tab:efimov_ratios}
    \begin{tabular}{@{}c c c c@{}}
        \toprule
        Ratio &
        Deterministic &
        DNN &
        DNN deviation from $e^{2\pi/s_0}$ (\%) \\
        \midrule
        $\epsilon_3^{(0)} / \epsilon_3^{(1)}$ & 507.75 & 507.87 & $-$1.39 \\
        $\epsilon_3^{(1)} / \epsilon_3^{(2)}$ & 515.02 & 500.54 & $-$2.81 \\
        $\epsilon_3^{(2)} / \epsilon_3^{(3)}$ & 515.03 & 512.70 & $-$0.45 \\
        \bottomrule
    \end{tabular}
\end{table}

\begin{figure}
    \centering
    \includegraphics[width=\columnwidth]{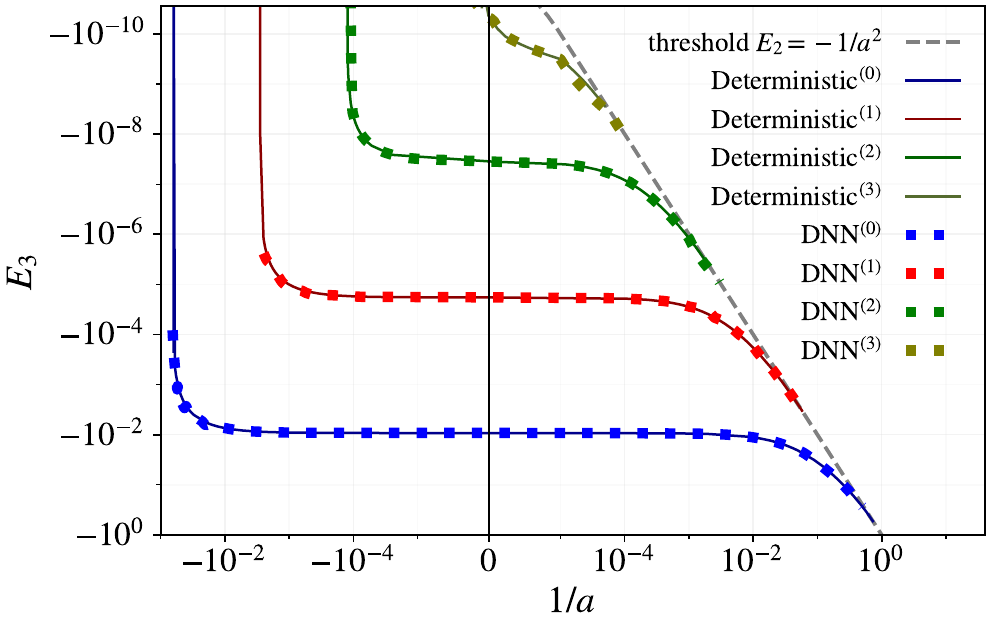}
    \caption{%
        Efimov spectrum $E_3$ as a function of the inverse scattering length
        for the first four three-body bound-state branches.
        Solid curves denote the deterministic solution obtained by diagonalizing
        the discretized subtracted kernel.
        Dotted curves denote the DNN continuation with trainable energy.
        The gray dashed line marks the atom-dimer threshold
        $E_2 = -1/a^2$ in the units used here ($\hbar = m = \mu_3 = 1$).
        The unitary limit is indicated by the vertical black line.
    }
    \label{fig:efimov_spectrum}
\end{figure}

Figure~\ref{fig:efimov_spectrum} shows the three-body energy $E_3$ as a function of the inverse scattering length for all four branches. 
The physical energy is negative ($E_3 = -\epsilon_3 < 0$) and is plotted on a symmetric-logarithmic scale to accommodate the wide range of binding energies. 
Solid curves denote the deterministic solution obtained by diagonalizing the discretized kernel at each $1/a$. 
Dotted curves denote the DNN with trainable energy, continued from the unitarity solution toward both negative and positive $1/a$. 
The gray dashed curve marks the atom-dimer threshold $E_2 = -1/a^2$, which is the boundary below which the two-body subsystem binds. 
On the positive scattering length side, the three-body branches approach this threshold from below, as expected for Efimov physics. 
The vertical black line at $1/a = 0$ marks the unitary limit. 
The visual agreement between deterministic and DNN curves extends over several decades in both $E_3$ and scattering length, 
demonstrating that the continuation strategy successfully propagates the DNN solution across the Efimov spectrum.

\begin{figure}
    \centering
    \includegraphics[width=\columnwidth]{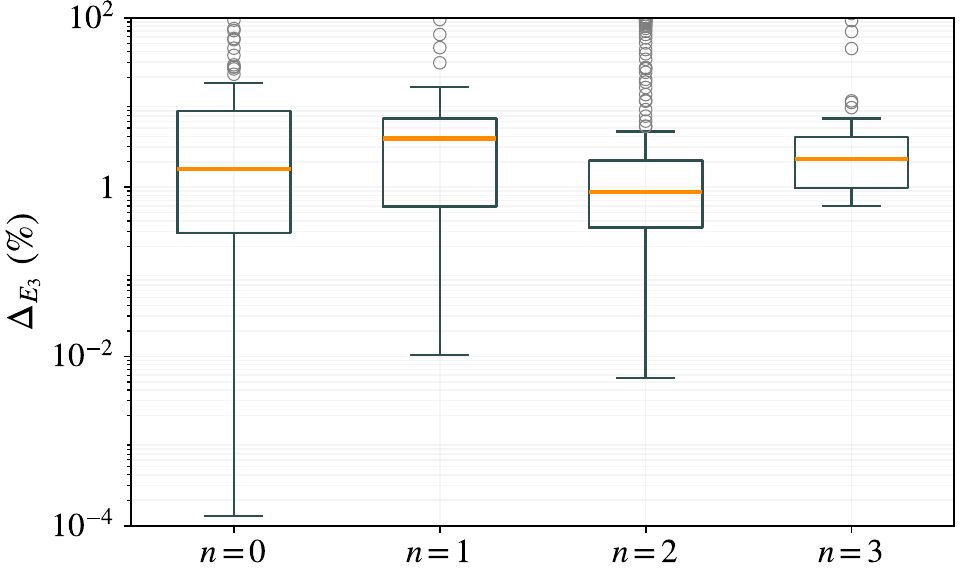}
    \caption{%
        DNN--deterministic relative deviation $\Delta_{E_3}$, Eq.~\eqref{eq:deltaE3},
        for the first four three-body bound states.
        The boxplot shows the distribution of $\Delta_{E_3}$ over all
        values of $1/a$ for which the DNN was continued.
        The deterministic benchmark is a numerical reference obtained from
        the same discretized kernel, not an exact analytical solution.
    }
    \label{fig:efimov_relative_deviation}
\end{figure}

To quantify the DNN--deterministic agreement across all values of $1/a$, we define the relative deviation
\begin{equation}\label{eq:deltaE3}
    \Delta_{E_3} =
    100\% \times \frac{|E_3^{\rm DNN}-E_3^{\rm det}|}{|E_3^{\rm det}|},
\end{equation}
where $E_3^{\rm det}$ is the deterministic energy obtained from the same discretized kernel. 
We emphasize that this quantity measures the consistency between two independent numerical 
strategies applied to the same integral equation, not an error relative to an exact analytical 
solution (which does not exist for arbitrary $1/a$). Figure~\ref{fig:efimov_relative_deviation} 
shows the distribution of $\Delta_{E_3}$ as a boxplot over all continuation points for each branch. 
The median deviations are $1.6\%$ ($n=0$), $3.8\%$ ($n=1$), $0.9\%$ ($n=2$), and $2.1\%$ ($n=3$). 
The lower quartiles lie below $1\%$ for all branches, while the medians remain at the percent level. 
The largest outliers appear predominantly on the positive-$1/a$ side near the atom--dimer threshold, 
where the two-body prefactor becomes nearly singular, and at large negative $1/a$, where the bound states 
become extremely shallow. These results indicate that the DNN provides a 
useful differentiable representation over the range studied, with reduced accuracy near thresholds and for very shallow branches.

\begin{figure}
    \centering
    \includegraphics[width=\columnwidth]{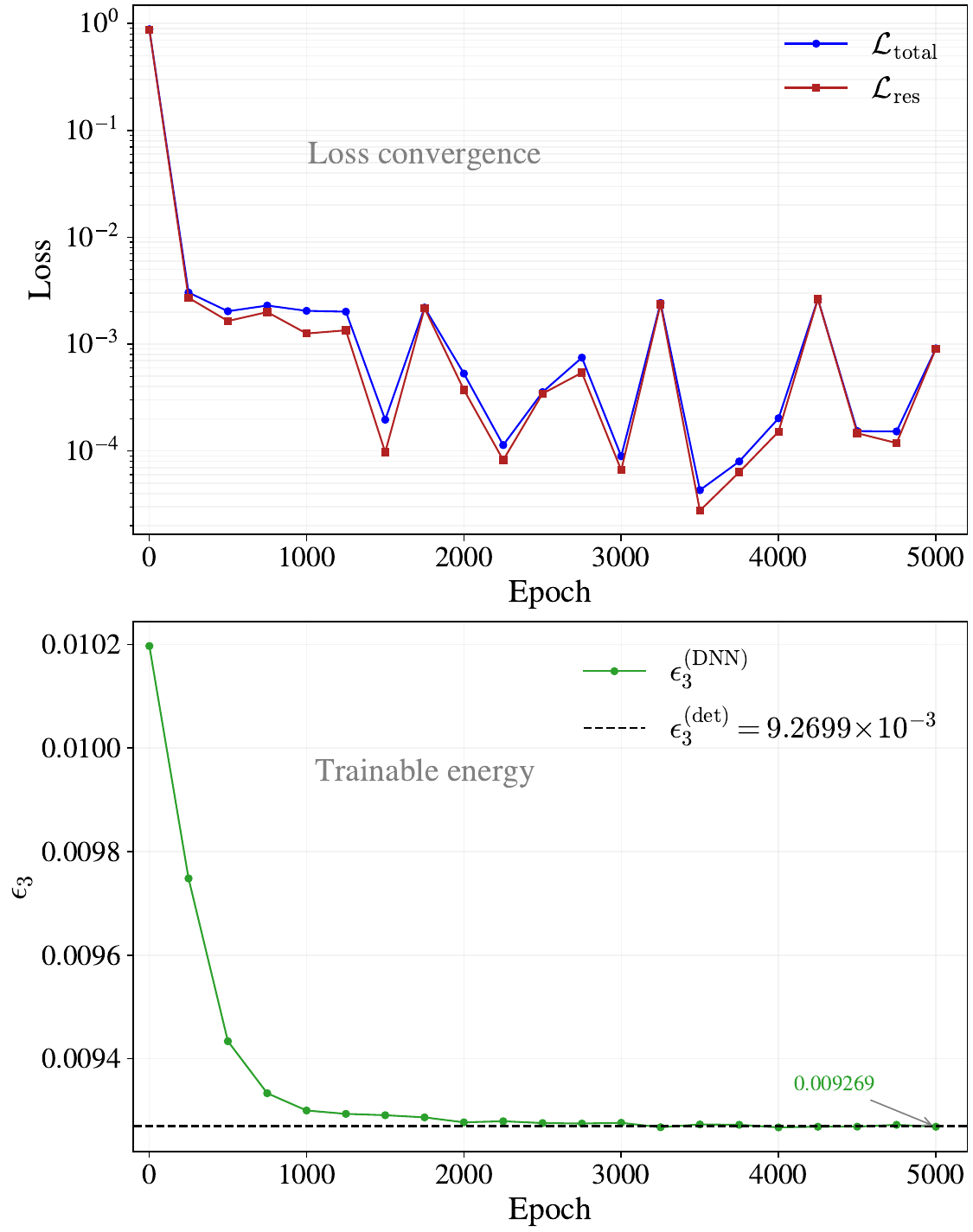}
    \caption{%
        Training convergence of the DNN solver for the Efimov ground state
        at unitarity.
        (a)~Residual loss $\mathcal{L}_{\rm res}$ (red squares) and total
        loss $\mathcal{L}_{\rm total}$ (blue circles) vs.\ epoch.
        (b)~Trainable binding-energy parameter $\epsilon_3^{\rm (DNN)}$
        vs.\ epoch; the dashed line marks the deterministic reference
        $\epsilon_3^{\rm (det)} = 9.2699\times10^{-3}$. 
    }
    \label{fig:efimov_gs_convergence}
\end{figure}

Finally, Fig.~\ref{fig:efimov_gs_convergence} shows the training trajectory for the ground state at unitarity, 
which serves as the starting point for all continuations. Panel (a) displays the residual loss 
$\mathcal{L}_{\rm res}$ and the total loss on a logarithmic scale: both decrease by roughly four orders 
of magnitude during the first 1500 epochs and reach values in the range $10^{-4}$--$10^{-5}$, with a 
best residual of $2.8\times10^{-5}$. Panel (b) tracks the trainable energy parameter $\epsilon_3^{\rm (DNN)}$, 
which starts $\sim 10\%$ above the deterministic value and converges to $\epsilon_3^{\rm (DNN)} = 9.2719\times10^{-3}$, 
within $0.022\%$ of $\epsilon_3^{\rm (det)}$. This convergence behavior validates the trainable-energy scheme and 
justifies using the unitarity solution as the initial condition for the continuation along $1/a$.

\section{Discussion}
\label{sec:discussion}

The results of Sec.~\ref{sec:results_efimov}  show that a DNN can be trained directly from the residual of a subtracted Faddeev integral equation, with the binding scale optimized as part of the same problem. This is the main methodological point of the calculation. The network is not fitted to deterministic eigenvectors or to precomputed energies; instead, it minimizes the mismatch of the discretized homogeneous equation while the kernel itself changes with the trainable parameter $\epsilon_3$. The agreement with the deterministic benchmark at unitarity and along the continuation in $1/a$ indicates that this residual formulation is sufficient to recover the relevant Efimov branches in the discretized problem.

The comparison with deterministic diagonalization should be interpreted as a consistency test between two numerical strategies for the same kernel. In the present one-dimensional zero-range problem, diagonalization and bisection remain the most direct way to obtain isolated bound states. The DNN formulation becomes interesting for a different reason: it produces a smooth, differentiable representation of the spectator amplitude and allows the solution at one point in parameter space to initialize the next one. This continuation property is particularly natural in problems where one wants to follow amplitudes as functions of scattering length, interaction parameters, or external constraints. Regarding computational cost, the present implementation rebuilds the $N_q \times N_q$ kernel matrix $B$ at every training epoch because $\epsilon_3$ changes, leading to a per-epoch matrix-vector cost of $O(N_q^2)$. This is not competitive with a single diagonalization for isolated bound states, but it is manageable for the grid sizes used here and does not dominate over the optimizer's own overhead.

The deviations observed for the shallowest branches and near the atom--dimer threshold are also informative. Efimov states span many orders of magnitude in binding energy, and the positive-$1/a$ side contains a near-singular two-body prefactor as the atom--dimer threshold is approached. In these regions, small residual mismatches can produce percent-level changes in the extracted binding scale. The DNN--deterministic deviations should therefore be read together with the residual history and the branch location in parameter space, rather than as a uniform accuracy estimate.

The present calculation is restricted to a single zero-range $s$-wave channel. Within this controlled setting, it establishes that equation-informed residual minimization can be applied to a renormalized few-body integral equation. Extensions to finite-range interactions, coupled spectator components, or continuum scattering observables are natural directions, but each would require significant modifications. Finite-range potentials introduce momentum-dependent form factors in the kernel. Coupled-channel problems demand a network with multiple output components and a matrix-valued residual that respects channel coupling symmetries. Scattering observables involve complex boundary conditions and require the network to represent non-normalizable continuum states, for which the homogeneous residual formulation used here is not directly applicable. Each of these directions therefore represents a nontrivial extension rather than a straightforward substitution. The differentiable-residual framework established here can serve as a starting point, but the required changes in both the kernel structure and the network architecture should not be underestimated.

\section{Conclusion}
\label{sec:conclusion}

We have presented a DNN residual-minimization approach for the subtracted three-body Faddeev spectator equation of identical 
bosons near the Efimov limit. The network represents the symmetrized spectator vector of the discretized integral equation, 
while the positive binding scale $\epsilon_3=-E_3$ is optimized jointly with the network weights. In this formulation, 
the DNN does not require an external spectral scan during training; the deterministic diagonalization of the same kernel is used only as an a posteriori benchmark.

At unitarity, the deterministic calculation recovers the expected Efimov scaling, while the DNN reproduces the two deepest binding scales with deviations of \(0.022\%\) and \(0.002\%\). 
The shallower branches are recovered at the percent level, consistent with their increased sensitivity to residual mismatch and grid resolution. Continuation in \(1/a\) 
then follows the deterministic branches across the range studied, with larger deviations localized near the atom--dimer threshold and in extremely shallow regions.
 
The main outcome of this proof-of-principle calculation is the construction of a differentiable DNN representation of the spectator 
amplitude, with the binding scale optimized as part of the residual minimization. 
For the present one-dimensional zero-range kernel, deterministic diagonalization remains the natural reference calculation. 
The DNN formulation provides a complementary representation: once trained, it provides a smooth continuation in $1/a$ and a compact 
differentiable amplitude that can be embedded in parameter scans, optimization loops, or future inverse formulations.

\acknowledgments
The author thanks Tobias Frederico for the encouragement to resume the study of few‑body systems using machine learning.
The code and data used to produce the results of this work are available from the author upon reasonable request.

\bibliographystyle{apsrev4-2}
\bibliography{references}

\end{document}